\documentclass[]{resources/aa}

\usepackage{graphicx}
\usepackage[varg]{txfonts}
\usepackage{physics}
\usepackage{amsmath, amssymb}
\usepackage{dsfont}
\usepackage{bbm}
\usepackage{xcolor}
\usepackage{ulem}
\usepackage{xpatch}
\usepackage{hyperref}
\usepackage{import}
\usepackage{makeidx}
\usepackage{changes}
\usepackage{footnote}

\usepackage{natbib}
\bibpunct{(}{)}{;}{a}{}{,} % to follow the A&A style

\newcommand{\sref}[1]{Sec.~\ref{#1}}
\newcommand{\tab}[1]{Table~\ref{#1}}
\newcommand{\fig}[1]{Fig.~\ref{#1}}
\newcommand{\equ}[1]{Eq.~\ref{#1}}
\newcommand{\equo}[1]{Eq.~\ref{#1}}
\newcommand{\equs}[2]{Eqs.~\ref{#1}~-~\ref{#2}}

\newcommand{\colout}[1]{\bgroup\markoverwith{\textcolor{#1}{\rule[.5ex]{2pt}{0.4pt}}}\ULon}

\makeatletter
\renewcommand*\aa@pageof{, page \thepage{} of \pageref*{LastPage}}
\newcommand{\pder}[2][]{\frac{\partial#1}{\partial#2}}
\makeatother

\makeindex

\begin{document}

\titlerunning{The importance of stellar rotation for protoplanetary disk evolution}
\authorrunning{L. Gehrig et al.}
\title{Time-dependent, long-term hydrodynamic simulations of the inner protoplanetary disk II: The importance of stellar rotation} 
% \subtitle{I. Modelling stellar period variability and its back reaction onto the disk}
%
% \subtitle{....}
%
\author{
 L.~Gehrig\inst{1},
 D.~Steiner\inst{1},
 E.~I.~Vorobyov\inst{1,2},
 M.~Güdel\inst{1}
}
\institute{
 Department of Astrophysics, University of Vienna,
 Türkenschanzstrasse 17, A-1180 Vienna, Austria
  \and
 Institute of Astronomy, Russian Academy of Sciences, 48 Pyatnitskaya St., Moscow, 119017, Russia
}
%\today
\date{Preprint online version: August 18, 2022}

\abstract
{The spin evolution of young protostars, surrounded by an accretion disk, still poses problems for observations and theoretical models. In recent studies, the importance of the magnetic star-disk interaction for stellar spin evolution has been elaborated. The accretion disk in these studies, however, is only represented by a simplified model and important features are not considered.}
{A more realistic representation of the accretion disk is indispensable for a better understanding of the star-disk interaction and the stellar spin evolution. The aim of this study is to investigate the influence of a hydrodynamic disk evolution on the stellar rotational period and vice versa during the accretion phase.}
{We combined the implicit hydrodynamic TAPIR disk code with a stellar spin evolution model. The influence of stellar magnetic fields on the disk dynamics, the radial position of the inner disk radius, as well as the influence of stellar rotation on the disk were calculated self-consistently.}
{Within a defined parameter space, we can reproduce the majority of fast and slow rotating stars observed in young stellar clusters. Additionally, the back reaction of different stellar spin evolutionary tracks on the disk can be analyzed. Disks around fast rotating stars are located closer to the star. Consequently, the disk midplane temperature in the innermost disk region increases significantly compared to slow rotating stars. We can show the effects of stellar rotation on episodic accretion outbursts. The higher temperatures of disks around fast rotating stars result in more outbursts and a longer outbursting period over the disk lifetime.}
{The combination of a long-term hydrodynamic disk and a stellar spin evolution model allows the inclusion of previously unconsidered effects such as the back-reaction of stellar rotation on the long-term disk evolution and the occurrence of accretion outbursts. However, a wider parameter range has to be studied to further investigate these effects. Additionally, a possible interaction between our model and a more realistic stellar evolution code (e.g., the MESA code) can improve our understanding of the stellar spin evolution and its effects on the pre-main sequence star.}


\maketitle

\section{Introduction}
\label{sec:intro}

Observations of young pre-main sequence (PMS) stars show that the stellar rotation periods range between $\sim 1 - 10$~days, which is significantly lower than the break up speed \citep[e.g.,][]{Irwin09, Gallet13, Serna2021}. Usually, these stars are surrounded by a protoplanetary disk \citep[e.g.,][]{Collier93, Hartmann98} and are expected to spin up due to the accretion of disk material containing high angular momentum during the PMS phase, which usually lasts $\lesssim 10$~Myr \citep[e.g.,][]{Mamajek09}. Additionally, the contraction of the young star should reduce the rotational period during the first several mega-year. Throughout the PMS phase, observations of stellar rotation lead to the assumption that the stellar period remains approximately constant during the disk phase \citep[e.g.,][]{Edwards93, Bouvier93, Irwin09, Gallet13}, implying the interaction between the star and the disk results in efficient breaking mechanisms that remove angular momentum from the star \citep[e.g.,][]{Serna2021}.

Early studies by \cite{Koenigl91}, based on the model introduced by \cite{Ghosh79}, describe angular momentum transfer via the star-disk interaction taking into account strong stellar magnetic fields. Such magnetic fields with a dipole strength of several kG \citep[e.g.,][and references therein]{Krull09, Johnstone14} have been observed and the field lines can truncate the disk up to several stellar radii \citep[e.g.,][]{Shu94, Romanova09, hartmann16}. Along these field lines the star and the disk can exchange angular momentum and the stellar rotational period can be regulated by the accretion disk (disk-locking model).
In more recent studies \citep[][]{Matt05, Zanni13}, the disk-locking model is shown not to be efficient enough to actually counteract stellar contraction and disk accretion, which in turn would spin up the star. The region, in which the stellar magnetic field lines connect to the disk is too narrow to extract sufficient angular momentum to match the observed stellar rotation periods.

Alternative breaking mechanisms slowing down the star depend on the outflow of material from the star and (or) the disk. First, the so called "X-wind" \citep[][]{Shu94} extracts material and angular momentum directly from the disk (close to the co-rotation radius) before it can be accreted onto the star. The amount of angular momentum due to accretion and thus the stellar spin-up is reduced. To launch the "X-wind," no intrinsic, large scale disk magnetic field is required \citep[e.g.,][]{Ferreira00} and the field lines are all anchored at the stellar surface.
If a large scale magnetic disk field is present, another kind of material outflow can carry away angular momentum from the star. The nature of this outflow depends on the orientation of the stellar magnetic moment compared to the disk's magnetic moment \citep[][]{Ferreira00, Romanova09, Zanni13}. 
In the case, where the stellar and disk's magnetic moment are aligned, the so called "ReX-wind" can be launched \citep[][]{Ferreira00}. At the point inside the disk, the so-called "X-point," where the stellar poloidal field cancels out the disk field, open field lines can reconnect to the closed stellar magnetic field. Matter parameterized by a specific fraction of the disk's accretion rate in \cite{Ferreira00} can be lifted onto and accelerated by stellar rotation along the open field lines at the "X-point" and thus removing angular momentum from the star. 
Magnetospheric ejections (MEs) can arise in case of a stellar and disk magnetic moment, which are anti-parallel \citep[][]{Romanova09, Zanni13}. A reoccurring process of inflation, opening, and re-connection of field lines leads to an out-bursting behavior, which can transfer angular momentum between the star and the disk.
Another model to explain the removal of angular momentum from the star is the accretion powered stellar wind (APSW) \citep[][]{Matt05b, Matt05, Matt08, Matt08b, Matt12, Reville15, Finnley18}. The wind is powered by a specific fraction of the disk's accretion rate and accelerated along the open stellar field lines by stellar rotation removing angular momentum from the star. 

An important parameter for the aforementioned theories is the disk's accretion rate \citep[e.g.,][]{Matt08b}. A precise calculation of the accretion rate including, for example, the influence of stellar magnetic torques and a back-reaction from the influence of stellar rotation on the disk appears rather difficult. Numerical time-step limitations, due to the Courant-Friedrichs-Lewy (CFL) condition \citep[][]{Courant28}, in the inner regions of the disk prevent the combined, self-consistent simulation of the stellar rotational evolution and hydrodynamic equations describing the protoplanetary disk. To circumvent this problem, previous studies have parameterized the disk's accretion rate by a simple power law or exponential decrease with respect to the disk lifetime \citep[e.g.,][]{Matt10, Gallet19}. An earlier attempt to combine disk and stellar spin evolution by \cite{Armitage96} has used the diffusion equation to simulate the disk.
While the disk's surface density can be evolved over time, assumptions have to be made, for example, concerning the angular velocity, which has to be made strictly proportional to the Keplerian angular velocity. Thus, a self-consistent calculation, especially of the inner disk parts is not possible as magnetic torques and local pressure gradient can alter the angular disk velocity \citep[e.g.,][]{Romanova02, bessolaz08, Steiner21} and consequently the accretion process itself.

In this study, we want to continue the work of the first paper of this series. \cite{Steiner21} have shown the importance of stellar magnetic torques for the long-term evolution of protoplanetary disks. We combine hydrodynamic disk and protostellar spin evolution to study the long-term effects of stellar rotation on the protoplanetary disk. The adaptive, implicit hydrodynamic (TAPIR) code for protoplanetary disks \citep[][]{ragossnig20, Steiner21}, which is based on \cite{Dorfi1998} and \cite{stoekl14a}, is extended with a stellar evolution model. Following \cite{Matt10}, this stellar evolution model includes a simplified description of the protostellar contraction (Kelvin-Helmholtz contraction). Additionally, the stellar spin evolution includes contributions from disk matter accreted onto the star, an accretion powered stellar wind (APSW), as well as magnetospheric ejections (MEs) \citep[e.g.,][]{Zanni13, Gallet19}.

By using a hydrodynamic rather than a diffusion approach, a self-consistent calculation of the innermost disk region is possible. This includes, for example, the radial position of the magnetic truncation radius, stellar magnetic torques, the influence of local pressure gradients, as well as episodic accretion outbursts. 
Those outbursts can be triggered by gravitational fragmentation in the disk outer parts followed by tidal destruction and accretion of inward-migrating clumps \citep[e.g.,][]{Vorobyov15}, by thermal instabilities \citep[TI,][]{bell94}, magneto-rotational instabilities (MRI) \citep[][]{Balbus1991, Armitage01} or a joint interaction of gravitational instability and MRI \citep[][]{Martin2011, Martin2012, Martin2013, Bae14}. In this study we focus on the occurrence of MRI triggered outbursts. We refer to the first paper of the series \citep[][]{Steiner21} for further details. Additionally, the effect of different stellar rotation periods on the disk's structure and evolution can be studied.

Due to the implicit nature of the TAPIR code, we can bypass the CFL condition and evolve our model for several mega-year \citep[e.g.,][]{Dorfi1998}.
Similar to \cite{Gallet19} and \cite{Steiner21}, we focus on Class~II star-disk systems, starting our simulations after the cloud infall has ended and the disk mass corresponds to $\lesssim 10$~\% of the stellar mass. Following \cite{Gallet19}, we choose an age of $t_\mathrm{0} = 1$~Myr as a start time for our simulations. 

Finally, we note that "X-winds" and "ReX-winds" are not modeled in this study. A suitable expression for "X-winds" is currently not available. The formulation for "ReX-winds" presented in \cite{Ferreira00} is very similar to the torque exerted by APSW and we follow \cite{Gallet19} including APSW and omitting "ReX-winds" in our model.

\section{Model description}
\label{sec:model_description}

In this chapter, the model used in our simulations is outlined. The focus of this study is the combination of a full hydrodynamic disk simulation and a stellar spin evolution model. The disk evolution is treated with the implicit hydrodynamics TAPIR code \citep[][]{ragossnig20, Steiner21} and the stellar spin model is formulated in detail in \cite{Matt10, Matt12} and \cite{Gallet19}. Here, we want to give a recap of the respective model's key features. In \sref{sec:hydrodynamic_disk_evolution}, \sref{sec:inner_disk_boundary} and \sref{sec:boundary_conditions} some key features of the TAPIR code, the treatment of the inner disk boundary are shown and the boundary conditions are outlined, respectively. The stellar spin model is depicted in \sref{sec:stellar_spin_model}.

\subsection{Hydrodynamic disk evolution with the TAPIR code}\label{sec:hydrodynamic_disk_evolution}

Contrary to long-term global disk simulations based on the diffusion evolution approach for the surface density $\Sigma$ \citep[e.g.,][]{pringle81, Armitage01, zhu07, zhu10a}, the TAPIR code allows the treatment of hydrodynamic equations. This includes a deviation from the Keplerian angular velocity $u_\mathrm{\varphi} \propto r^{-1/2}$. Compared to the model presented in \cite{ragossnig20}, we added the influence of the stellar magnetic field, which is especially important in the inner part of the disk \citep[e.g.,][]{Romanova02, bessolaz08}. 
The disk is assumed to be axisymmetric in angular direction and to be in hydrostatic equilibrium in vertical direction.
Our model is formulated in cylindrical coordinates with the planar components (r,$\varphi$), which are then discretized \citep[for a comprehensive description regarding discretization and numerical properties, see][]{ragossnig20}.
The protoplanetary disk in our simulations is described in \equs{eq:cont}{eq:ene} as a time-dependent, vertically integrated, viscous accretion disk \citep[e.g.,][]{shakura73, Armitage01},
\begin{alignat}{2}
    & \pder{t} \, \Sigma &&+ \nabla \cdot ( \Sigma \, \vec u ) = 0\;, \label{eq:cont} \\
    & \pder{t} (\Sigma \, \vec u) &&+ \nabla \cdot (\Sigma \, \vec u : \vec u) - \frac{B_\mathrm{z} \vec B}{2 \pi}\nonumber \\
    & &&+ \nabla P_\mathrm{gas} + \nabla \cdot Q + \Sigma \, \nabla \psi + H_\mathrm{P} \nabla \left( \frac{B_\mathrm{z}^2}{4 \pi} \right) = 0 \;, \label{eq:mot} \\
    &\pder{t} (\Sigma \, e) &&+ \nabla \cdot (\Sigma \, \vec u \, e ) + P_\mathrm{gas} \, \nabla \cdot  \vec u \nonumber \\
    & &&+ Q : \nabla \vec u - 4 \pi \, \Sigma \, \kappa_\mathrm{R}\left(J - S \right) + \dot E_\mathrm{rad} = 0 \;, \label{eq:ene}
\end{alignat}
where $\Sigma$, $\vec u$, $e$, $P_\mathrm{gas}$ and $H_\mathrm{P}$ denote the gas column density, gas velocity in the planar  components $\vec u = (u_\mathrm{r}, u_\mathrm{\varphi})$, specific internal energy density, vertically integrated gas pressure and the vertical scale height of the gas disk, respectively.
The gradient in planar cylindrical coordinates reads $\nabla = (\partial / \partial r , r^{-1} \partial / \partial \varphi)$ with $\partial / \partial \varphi = 0$ for our axisymmetric model.
$P_\mathrm{gas}$ is utilized by the ideal equation of state $P_\mathrm{gas} = \Sigma e ( \gamma - 1 )$, with the adiabatic coefficient $\gamma = 5 / 3$ \citep[e.g.,][]{bessolaz08} and $\kappa_\mathrm{R}$ denotes the Rosseland-mean opacity \citep[e.g.,][]{Mihalas84}, which is composed of a gaseous component $\kappa_\mathrm{R,gas}$ \citep[based on][]{caffau11} and a dust-dominated component $\kappa_\mathrm{R,dust}$ \citep[based on][]{pollack85}; with $\kappa_\mathrm{R} = \kappa_\mathrm{R,gas} + f_\mathrm{dust} \kappa_\mathrm{R,dust}$, where $f_\mathrm{dust} = 0.01$ is the dust-to-gas mass ratio.
The gravitational potential of the star-disk system is denoted by $\psi$.
Additionally, the effects of a stellar magnetic field on the disk are included. The vertical component of the magnetic field $B_\mathrm{z}$ is assumed to be a dipole and constant within the disk's vertical extent but can change radially. The planar magnetic field components $\vec B = (B_\mathrm{r}, B_\mathrm{\varphi})$ are taken at the surface of the disk. We note that we neglect the radial component of the magnetic field in this study ($B_\mathrm{r} = 0$, see \citealt{rappaport04} or \citealt{kluzniak07}).  The last term on the left-hand side of \equ{eq:mot} is the vertically integrated magnetic pressure gradient and $B_\mathrm{z} \vec B$ denotes the magnetic stress induced by a Lorentz force \citep[as in][]{Vorobyov20}, which exerts a magnetic torque per unit area on the disk $\sim r B_\mathrm{z} \vec B$ \citep[e.g.,][]{Lubow94}. 
The field is assumed to be in rigid rotation with the central star and is able to exchange angular momentum between the star and the disk (\sref{sec:stellar_spin_model}). Viscous angular momentum transport and viscous heating are represented by the viscous stress tensor $Q$ in their corresponding terms in \equ{eq:mot} and \equ{eq:ene}, respectively, which reads as,
\begin{align}
    &Q = \frac{\mu}{2} \left[ \nabla \vec{u} + (\nabla \vec{u})^T - \frac{2}{3}(\nabla \cdot \vec{u})~\mathds{1} \right] \;. \label{eq:shearstresstensor}
\end{align}
The factor $\mu$ denotes the vertically integrated dynamical viscosity, which yields,
\begin{align}
    \nu &= \alpha \, c_\mathrm{S} \,  H_\mathrm{P} \;, \label{eq:shakura} \\
    \mu &=  \nu \, \Sigma  \;, \label{eq:mu}
\end{align}
where $\alpha$, $c_\mathrm{s}$, $H_\mathrm{p}$ and $\nu$ depict the viscosity parameter \citep[][]{Shakura1973}, the speed of sound, the vertical pressure scale height and the kinematic viscosity, respectively. 

The disk is irradiated by the star and cools radiatively, which is described by the net stellar heating and cooling rate per unit surface area $\dot E_\mathrm{rad}$. 
$\dot E_\mathrm{rad}$ is obtained by balancing stellar irradiation, radiative cooling at the disk surface and irradiation from an ambient medium \citep[see][]{Steiner21}.
The $(J - S)$-term in \equ{eq:ene} depicts radial radiative transport and is modeled in a radiative diffusion approximation with an Eddington factor of $1/3$ \citep[e.g.,][]{ragossnig20, Steiner21}.
We note that disk heating due to magnetic coupling to the star is not included in \equ{eq:ene}. 
Work done by the magnetic star-disk interaction is assumed to be comparable to $P\Delta V$~work only close to the inner disk edge as the stellar field strength decreases with $r^{-3}$ (see \sref{sec:inner_disk_boundary}). 
In this region close to the star, however, the disk temperature is dominated by stellar irradiation. 
If large scale disk fields are taken into account, disk heating due to magnetic fields can affect the disk's structure and evolution and is added to \equ{eq:ene} in future studies (Steiner~et~al.~in~prep).
Furthermore, resistive dissipation in the disk is assumed to be negligible compared to $P\Delta V$~work \citep[e.g.,][]{Miller1997}.
Additionally, we utilized the adaptive grid formulation introduced by \cite{Dorfi1987} to provide adequate radial grid point resolution and allow the inner radius to move according to the disk evolution (see \sref{sec:inner_disk_boundary}). The grid is solved together with \equs{eq:cont}{eq:ene} to prevent artificial numerical oscillations \citep[e.g.,][]{Dorfi1987, ragossnig20} from developing.

\subsection{Inner disk boundary}\label{sec:inner_disk_boundary}

In the innermost disk region, inside the co-rotation radius the stellar magnetic field decelerates the disk material. As the disk's angular velocity $u_\mathrm{\varphi}$ becomes sub-Keplerian, the disk material accelerates radially toward the star  \citep[e.g.,][]{Romanova02, Steiner21}. This further leads to a drop in the surface density $\Sigma$ \citep[e.g.,][]{bessolaz08, Steiner21} toward the star and strong localized pressure gradients $d P_\mathrm{gas} / dr$. The stellar magnetic torque as well as pressure gradients imply a significant deviation from the commonly used 1D diffusion equation for describing the long-term evolution of $\Sigma$ \citep[e.g.,][]{bell94, Armitage01, zhu09}, therefore it is necessary to fully solve the equations of hydrodynamics.
The disk is assumed to eventually be disrupted by a strong stellar magnetic field at the magnetic truncation radius $r_\mathrm{trunc}$. Inside $r_\mathrm{trunc}$ the gas does not behave as an accretion disk but flows along magnetic funnels toward the star \citep[e.g.,][]{Romanova02, Romanova14}. This magnetically dominated region cannot be described by a 1D model, which justifies the choice of our inner boundary $r_\mathrm{in}$ to be at $r_\mathrm{in} = r_\mathrm{trunc}$. The stellar magnetic field $\vec B$ is approximated as a dipole field being in corotation with the star. The vertical component $B_\mathrm{z}$ and angular component $ B_\mathrm{\varphi}$ of $\vec B$  are modeled as \citep[e.g.,][]{rappaport04, kluzniak07}
\begin{alignat}{2}
    & B_\mathrm{z}(r) &&= B_\star \left(\frac{R_\star}{r}\right)^3  \label{eq:stellar_Bz} \; \mathrm{and}\\
    & B_\mathrm{\varphi}(r) &&\simeq - \alpha_\mathrm{cor} \, B_\mathrm{z}(r) \left[ 1- \left(\frac{\Omega_\star}{\Omega(r)}\right)^{\alpha_\mathrm{cor}} \right] \;, \label{eq:stellar_Bphi}
\end{alignat}

with 

\begin{alignat}{2}
      & \alpha_\mathrm{cor} &&\equiv
      \begin{cases}
      +1 & \text{if $r < r_\mathrm{cor}$} \\
      -1 & \text{if $r > r_\mathrm{cor}$} \, , \\
    \end{cases} \\
\end{alignat}

\noindent as well as the co-rotation radius

\begin{alignat}{2}
    &r_\mathrm{cor} &&= \left( \frac{G \, M_\star \, P_\star^2}{4 \, \pi^2} \right)^{1/3} \label{eq:corotation_radius} \;.
\end{alignat}
Here, $B_\star$, $M_\star$, $\Omega_\star$, $P_\star$ and $G$ denote the magnetic field strength at the stellar surface, the stellar mass, the stellar angular velocity, the stellar rotation period and Newton’s gravitational constant, respectively.  The factor $\alpha_\mathrm{cor}$ switches sign at the corotation radius $r_\mathrm{cor}$ and accounts for the braking region inside $r_\mathrm{cor}$ (field winds up as a trailing spiral) and the propeling region outside $r_\mathrm{cor}$ (field winds up as a leading spiral). Additionally, $\alpha_\mathrm{cor}$ ensures that $|B_\mathrm{\varphi}| \simeq B_\mathrm{z}$ for all radii $r \ll r_\mathrm{cor}$ and $r \gg r_\mathrm{cor}$, which is a physically motivated model, since a tightly wound up field leads to open field lines due to magnetic buoyancy and therefore limits the maximum value of $|B_\mathrm{\varphi}|$ \citep[e.g.,][]{Lynden-Bell1994, rappaport04}.\\

The magnetic truncation radius $r_\mathrm{trunc}$ denotes the point where the magnetic pressure $P_\mathrm{mag}$ equals either the ram pressure $P_\mathrm{ram}$ of the in-falling material or the gas pressure $P_\mathrm{gas}$, depending on which pressure contribution dominates \citep[e.g.,][]{Koldoba02, Romanova02, bessolaz08},
\begin{equation}\label{eq:r_trunc}
    P_\mathrm{magn}(r_\mathrm{trunc}) = \text{max}\left[P_\mathrm{ram}(r_\mathrm{trunc}), P_\mathrm{gas}(r_\mathrm{trunc})\right]\;.
\end{equation}
This yields for $r_\mathrm{in} = r_\mathrm{trunc}$ the following description for the case $P_\mathrm{ram} \geq P_\mathrm{gas}$ \citep[][]{hartmann16},
\begin{align}
    r_{\mathrm{trunc}}(P_\mathrm{ram} \geq P_\mathrm{gas}) \approx 18 \, \xi \, R_\odot & \, \left(\frac{B_\star}{10^3 \, G}\right)^{4/7}  \left(\frac{R_\star}{2 \, R_\odot}\right)^{12/7} \left(\frac{M_\star}{0.5 \, M_\odot}\right)^{-1/7} \nonumber \\ 
    & \left(\frac{\dot M_\star}{10^{-8} \, M_\odot / \mathrm{yr}}\right)^{-2/7} \label{eq:truncation_radius} \;, 
\end{align}
with the accretion rate onto the star $\Dot{M}_\star$ and a correction factor $\xi = 0.7$ \citep[following][]{hartmann16} accounting for non-spherical accretion and the
% complicated
oversimplified star-disk interaction. The case $P_\mathrm{ram} < P_\mathrm{gas}$ follows from equating the magnetic pressure of the vertical field component $B_\mathrm{z}$ (\equ{eq:stellar_Bz}) with the gas pressure,
\begin{align}
    &r_{\mathrm{trunc}}(P_\mathrm{ram} < P_\mathrm{gas}) = \frac{{B_\star}^{1/3} \, R_\star}{{P_\mathrm{gas}}^{1/6} \, (8 \pi )^{1/6} } \;. \label{eq:truncation_radius_press}
\end{align}

\subsection{Numerical boundary conditions}\label{sec:boundary_conditions}

The numerical boundary conditions for the hydrodynamical quantities used in the TAPIR code (cp.~\equs{eq:cont}{eq:ene}) are chosen according to \cite{Romanova02, Romanova04} as "free" (Neumann) boundary conditions (e.g., $\partial \Sigma / \partial r = 0$) at the inner boundary. 
At the outer boundary, the surface density and the internal energy are also treated with "free" boundary conditions, while the angular velocity $u_\mathrm{\varphi}$ is fixed to the Keplerian value and radial velocity $u_\mathrm{r}$ is set to zero representing no inflow of material from an parent cloud or outer envelope, as usual in late Class~II systems.
The position of the outer boundary is set to the position where $\Sigma = 0.5$~$\mathrm{g/cm^2}$ \citep[similar to][]{Vorobyov20} and is allowed to move during the simulation according to the disk evolution.

\subsection{Stellar spin model}\label{sec:stellar_spin_model}

The spin evolution model in this work considers the torques acting on the young star from external contributors as well as the internal stellar evolution. 
The temporal derivative of angular momentum equals the sum of external torques $\Dot{J}_\star = \Gamma_\mathrm{ext}$. Following \cite{Matt10}, we assume rigid body rotation with a moment of inertia $I_\star = k^2 M_\star R_\star^2$ with the normalized radius of gyration $k$ \citep[$k^2 = 0.2$, e.g.,][]{Armitage96, Matt10, Matt12}.
The temporal derivative of the stellar angular momentum $J_\star = I_\star \Omega_\star$ reads
\begin{equation}\label{eq:stellar_angular_momentum}
    \Dot{J}_\star = I_\star \Dot{\Omega}_\star + \Dot{I}_\star \Omega_\star = \Gamma_\mathrm{ext} \, .
\end{equation}
The evolution of the stellar spin results from rewriting \equ{eq:stellar_angular_momentum} as follows
\begin{equation}\label{eq:stellen_spin_evolution}
    \Dot{\Omega}_\star = \frac{\Gamma_\mathrm{ext}}{I_\star} - \frac{\Dot{I}_\star}{I_\star} \Omega_\star \, ,
\end{equation}
with 
\begin{equation}\label{eq:dotI_I}
    \frac{\Dot{I}_\star}{I_\star} = \frac{\Dot{M}_\star}{M_\star} + 2 \frac{\Dot{R}_\star}{R_\star} \, .
\end{equation}

Following the model of \cite{Gallet19}, the external torque $\Gamma_\mathrm{ext}$ acting on the star consists of three components; $\Gamma_\mathrm{ext} = \Gamma_\mathrm{acc} + \Gamma_\mathrm{w} + \Gamma_\mathrm{ME}$.
The torques resulting from accreting disk material (gas and dust) onto the star $\Gamma_\mathrm{acc}$, stellar winds and outflows $\Gamma_\mathrm{w}$ and magnetospheric ejections $\Gamma_\mathrm{ME}$ are described in this section. 
As we focus on young class II disk systems, we assume a fully convective star. Thus, internal torques resulting from the interaction between a radiative core and a convective envelope \citep[as describe in][]{Gallet19} are not considered in this work. 
The change of the stellar moment of inertia due to contraction of the stellar radius is included in our model.

\subsubsection{Stellar radius evolution}\label{sec:stellar_radius_evolution}

We assume a fully convective young star \citep[e.g.,][]{Matt10, Pantolmos20} contracting during the pre main-sequence until a radiative core develops \citep[][]{Siess00}. 
During this phase, the stellar radius evolution can be approximated by a Kelvin-Helmholtz contraction \citep[e.g.,][]{Collier93, Matt10, Matt12} resulting in

\begin{equation}\label{eq:stellar_radius_evolution}
    \Dot{R}_\star = 2 \frac{R_\star}{M_\star} \Dot{M}_\star - \frac{28 \pi \sigma R_\star^4 T_\mathrm{eff}^4}{3 \mathrm{G} M_\star^2} \, ,
\end{equation}
with the stellar effective temperature $T_\mathrm{eff}$ and the Stefan-Boltzmann constant $\sigma$. The first term on the left hand side of \equ{eq:stellar_radius_evolution} accounts for accreted disk material. The second term represents the contraction on the Kelvin-Helmholtz timescale. 

We set the initial stellar mass of the star to $M_\star = 0.7~\mathrm{M_\odot}$. A broader range of stellar masses are included into future studies. The respective $R_\mathrm{\star,~init}$ and $T_\mathrm{eff,~init}$ are taken from the isochrones of \cite{Baraffe15}. The stellar radius then evolves according to \equ{eq:stellar_radius_evolution} and the effective temperature updates according to the results of \cite{Baraffe15} with time.
We want to note that this stellar evolution model is a simple approximation as several important aspects of the pre-main-sequence evolution are ignored \citep[e.g., Deuterium burning slows down the stellar contraction;][]{Siess00, Matt10}. The maximum difference between \equ{eq:stellar_radius_evolution} and the results in \cite{Baraffe15} is $<2.5$~\% for a 0.7~$M_\odot$ star during the first 10~Myr of its lifetime, assuming no accretion.

\subsubsection{Magnetospheric ejections}
\label{sec:ME}

In star-disk systems, with a stellar magnetic field strong enough to connect to the disk material, angular momentum can be transferred between the star and the disk by magnetospheric ejection (ME) \citep[][]{Zanni13}, which is nicely summarized in \cite{Gallet13}. Differential rotation between the star and the disk generates a toroidal magnetic field. The resulting magnetic pressure inflates the magnetic field lines, followed by a re-connection and contraction \citep[see Fig.~2 of][]{Zanni13}.
The magnetic torque felt by an annulus ($\Delta r$) of the disk \citep[e.g.,][]{Armitage96, Matt05b, Gallet19, Ireland21} is given by 
\begin{equation}\label{eq:gamma_me_annulus}
    \Delta \Gamma_\mathrm{ME} = q r^2 B_\mathrm{z}^2 \Delta r \, ,
\end{equation}
with the twist of magnetic field lines  
\begin{equation}
 q \propto \frac{B_\mathrm{\varphi} }{ B_\mathrm{z}} \sim \left[ 1 - \left( \frac{r_\mathrm{trunc}}{r_\mathrm{cor}} \right)^{3/2}  \right] \, .
\end{equation}

The additional assumption that the radius of the annulus of the disk connected to the MEs is of the order of $r_\mathrm{trunc}$ \citep[][]{Gallet19} yields the total torque felt by the star due to MEs, 
\begin{equation}\label{eq:gamma_me}
    \Gamma_\mathrm{ME} = K_\mathrm{ME} \frac{B_\star^2 R_\star^6}{r_\mathrm{trunc}^3} \left[ K_\mathrm{rot} - \left( \frac{r_\mathrm{trunc}}{r_\mathrm{cor}} \right)^{3/2} \right] \, ,
\end{equation}
with $K_\mathrm{ME} = 0.21$ and $K_\mathrm{rot} = 0.7$. This parameter choice represents the fact that the MEs rotate at a sub-Keplerian rate \citep[see][]{Gallet19}. We want to note that \cite{Pantolmos20} found smaller values for $K_\mathrm{ME} = 0.13$ and $K_\mathrm{rot}= 0.46$. 
The differences are explained by slightly different boundary definitions for the stellar wind and MEs. Furthermore, \cite{Ireland21} include multi-dimensional and non-linear effects to the calculation of $\Gamma_\mathrm{ME}$ and also derive lower values for the constant of proportionality compared to \cite{Gallet19}. 
In this study we use the values given by \cite{Gallet19}, which can be seen as upper limits for $\Gamma_\mathrm{ME}$.

The effect of the star-disk interaction depends on the position of the inner radius $r_\mathrm{trunc}$ in relation to the corotation radius $r_\mathrm{cor}$. If $r_\mathrm{trunc} < K_\mathrm{rot}^{2/3} r_\mathrm{cor}$, the stellar angular momentum increases and the star spins up, otherwise the star spins down. As argued in \cite{Zanni13}, MEs reduce the effect of the accretion torque $\Gamma_\mathrm{acc}$.
In \cite{Gallet19}, a constant $K_\mathrm{acc} < 1$ qualitatively represents this reduction of $\Gamma_\mathrm{acc}$ (see \sref{sec:acc}). $K_\mathrm{acc}$ serves as an approximation to quantify small-scale, time-dependent magnetically mediated mass transfer between the star and the disk. Following \cite{Gallet19}, we choose $K_\mathrm{acc} = 0.4$ for all our simulations.
As shown in \cite{Romanova09}, the process is very dynamic and thus cannot be adequately described in a 1D model. The following approximations should be seen as time averaged values. We refer the reader to \cite{Romanova09} and \cite{Zanni13} for a more elaborated view on the topic.

\subsubsection{Accretion}
\label{sec:acc}

During the accretion of material onto the young star, mass and angular momentum is added to the central object. The accretion torque $\Gamma_\mathrm{acc}$ is given by the product of the specific angular momentum of disk material at the inner edge of the disk and the accretion rate;

\begin{equation}
    \Gamma_\mathrm{acc} = K_{acc} \Dot{M}_\mathrm{\star} r_\mathrm{trunc}^2 \Omega_\mathrm{disk,in}
\end{equation}

with $r_\mathrm{trunc} \Omega_\mathrm{disk,in} = u_\mathrm{\varphi, in}$.

\subsubsection{Stellar outflow}
\label{sec:winds}

In a magnetized stellar outflow, material is transported along magnetic field lines. These field lines are rooted on the star and rotate with $\Omega_\star$. As the wind material is moving outwards, it magneto-centrifugally accelerates and thus removes angular momentum from the star. 

In this study, we consider the APSW \citep[][]{Matt05b}, which takes its driving force from the accreting disk material. Consequently, the mass loss due to this wind is a fraction $W$ of the disk accretion rate; $\Dot{M}_\mathrm{W} = W \Dot{M}_\mathrm{disk}$. 
The torque, caused by a stellar outflow \citep[][]{Weber67}, acting on the star, can be written as 
\begin{equation}\label{eq:gamma_wind}
    \Gamma_\mathrm{w} = \Dot{M}_\mathrm{w} r_\mathrm{A}^2 \Omega_\star  \, ,
\end{equation}
with the Alfv\'en radius 
\begin{equation}\label{eq:alfven}
    r_\mathrm{A} = K_\mathrm{1} \left[ \frac{B_\star^2 R_\star^2}{\Dot{M}_\mathrm{wind} \sqrt{K_\mathrm{2}^2 v_\mathrm{esc}^2 + \Omega_\star^2 R_\star^2} }  \right]^m R_\star \, ,
\end{equation}
where $v_\mathrm{esc} = \sqrt{2 \mathrm{G} M_\star / R_\star}$, $m = 0.2177$, $K_\mathrm{2} = 0.0506$ and $K_\mathrm{1}~=~1.7$ arising as best fit values from magneto-hydrodynamic stellar wind simulations \citep[see][]{Matt12b, Gallet19}. We want to note that $\Gamma_\mathrm{w}$ acts as an angular momentum "sink" in our model. The exact driving mechanisms of the APSW and the order of $W$ are still under debate (we refer to the discussion in \cite{Gallet19} for further details).

\subsubsection{Numerical integration}

To include the spin evolution model in the TAPIR code, we updated the stellar parameters after each timestep $\Delta t$ as follows. The current accretion rate resulting from solving the disk equation \equs{eq:cont}{eq:ene} updates the stellar mass and radius. With the updated values the respective torque contributions acting during this timestep are calculated according to the formulation above. Finally, $\Dot{\Omega}$ and the new stellar angular velocity $\Omega_\mathrm{new} = \Omega_\mathrm{old} + \Dot{\Omega} \Delta t$ is derived and used in the next timestep. $\Omega_\mathrm{old}$ and $\Omega_\mathrm{new}$ denote the stellar angular velocity
used for the current and the new timestep, respectively.

%% \import{sections/}{spin_model}

\section{Results}
\label{sec:results}

In this section, we want to define the parameter space and a reference model (\sref{sec:parameter_space}), followed by the results of our simulations: the back-reaction of different stellar rotation rates on the disk (\sref{sec:backreaction}), the construction of a fast and a slow rotating protostar within our parameter space (\sref{sec:fast_slow}), as well as the influence of stellar rotation on episodic accretion outburst events (\sref{sec:outbursts}).

\subsection{Parameter space}
\label{sec:parameter_space}

We define a reasonable parameter space for stellar and disk properties covered in our simulations, which is based on observations and previous theoretical models.
First we define the stellar parameters appropriate for a 1~Myr old T Tauri star. As described in \sref{sec:model_description}, the initial stellar mass is $M_\star = 0.7 \mathrm{M_\odot}$ and the initial stellar radius, as well as the effective temperature are taken from isochrones provided by \cite{Baraffe15}.
The initial stellar rotation period for a 1~Myr old star can vary significantly. We choose a range for the initial period from 2 to 15~days. This choice is motivated by observations of the Orion Nebular Cluster (ONC) \citep[][]{herbst02}. 
The stellar dipole magnetic field strength has to be estimated. Measurements by \cite{Johnstone14} suggest that the range of the dipole component of the magnetic field can range from several hundred Gauss to $\sim$~kG. 
Furthermore, M~dwarfs can have very large magnetic field strengths of several thousand Gauss \citep[e.g.,][]{Kochukhov20}.
Based on these measurements we assume $B_\star$ to be in a range between $500$ and $2000$~G.
The magnetic field strength remains constant during our simulations.
We note that T~Tauri stars are usually magnetically saturated \citep[Rossby number $\lesssim 0.1$, e.g.,][]{Briggs07, Johnstone2021} and the stellar magnetic field is roughly independent of the rotational period \citep[e.g.,][]{Reiners2009, Lavail2019}. During later evolutionary stages, this relation is reverted and fast rotating stars show more magnetic activity up to the state of magnetic saturation \citep[e.g.,][]{Wright11}.
Finally, the efficiency $W$ of the APSW has to be chosen. We choose three different values for $W$. Based on the theoretical model by \cite{Crammer08} and simulations by \cite{Pantolmos20}, we choose $W=2\%$ as a reference value. $W=0\%$ and $W=5\%$ represent a weak and strong APSW, respectively. 

The hydrodynamic disk model requires an initial disk accretion rate as well as description of the viscous $\alpha$ parameter. Based on observations \citep[e.g.,][]{Gullbring98,Manara16} and reproduced by numerical simulations \citep[e.g.,][]{Vorobyov17c}, the accretion rate from the disk onto the star ranges between $10^{-10}$ and $10^{-7}~\mathrm{M_\odot \, \mathrm{yr}^{-1}}$ for late-type star-disk systems. 
We therefore choose the initial accretion rate $\Dot{M}_\mathrm{init} = 1 \cdot 10^{-8}~\mathrm{M_\odot/yr}$. The effects of a variation of $\Dot{M}_\mathrm{init}$ are covered in a further study.
The viscous $\alpha$-parameter ranges between $\lesssim 0.1$ and $\sim 0.0001$ \citep[e.g.,][]{zhu07, zhu09b, Vorobyov09, Yang18}. We adopt initially a constant value of $\alpha = 0.005$ used in \sref{sec:backreaction} and \sref{sec:fast_slow}, unless otherwise stated.

To generate episodic accretion outbursts (\sref{sec:outbursts}) due  to magnetorotational instabilities (MRI) \citep[][]{Balbus1991},  we adapt the layered disk model \citep[][]{gammie96}. A detailed description of this model in the TAPIR code is presented in \cite{Steiner21}. In the layered disk model, the disk is divided in two layers. On the one hand, we assume an MRI active layer, which is sufficiently ionized due to stellar irradiation, cosmic rays or collisional ionization of, for example, potassium \citep[][]{gammie96} with an effective viscosity $\alpha_\mathrm{MRI}$. On the other hand, we have an MRI inactive layer, the dead zone (DZ), with $\alpha_\mathrm{DZ} \ll \alpha_\mathrm{MRI}$, in which the mass transport rate is strongly diminished. 
This leads to mass piling up in the DZ up to a point where the increasing mass leads to increasing temperature and eventually to onset of the MRI above a critical temperature $T_\mathrm{crit}$. 
The turbulent viscosity and viscous heating increase sharply, which further causes ionization of the whole inner region and a highly increased mass transport rate $\dot M$. This ionized inner region is then accreted in an strong accretion event onto the star and the inner disk is replenished by material from the outer disk, which completes a burst cycle.

\begin{figure}[ht]
    \centering
         \resizebox{\hsize}{!}{\includegraphics{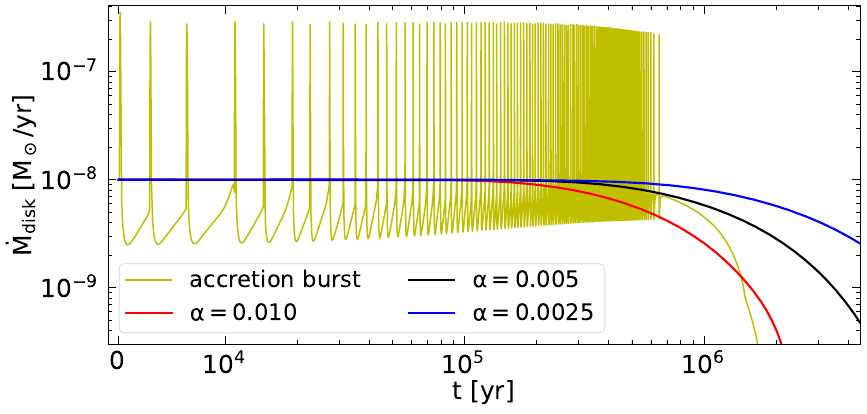}}
    \caption{Disk accretion rate $\Dot{M}_\mathrm{disk}$ with respect to the simulation time $t$ in years starting from $t_\mathrm{0} = 1$~Myr. The red, black, blue lines correspond to a constant $\alpha$ value of 0.01, 0.005 and 0.0025, respectively. The yellow line corresponds to a layered disk model \citep[][]{gammie96} with $\alpha_\mathrm{MRI} = 0.02$ and $\alpha_\mathrm{DZ} = 0.0004$ \citep[similar to][]{Steiner21}.}
    \label{fig:alpha}
\end{figure}

Depending on the choice of the previously mentioned parameters, the evolution of the accretion disk and therefore the spin evolution of the star can vary. One example is the viscosity parameter $\alpha$ that influences the viscous timescale of the accretion process; $\tau_\mathrm{\nu}=r^2/\nu\sim 1/\alpha$. A large~(small) $\alpha$-value corresponds to short~(long) disk lifetime, respectively, and the accretion rate changes accordingly \citep[e.g.,][]{Vorobyov09}. In our model, we can include these effects self-consistently. In \fig{fig:alpha}, the effect of different $\alpha$-values on the disk evolution is shown. The red, black and blue lines show the temporal evolution of the disk's accretion rate $\Dot{M}_\mathrm{disk}$ for a constant $\alpha$-value of 0.01, 0.005 and 0.0025, respectively. The time for $\Dot{M}_\mathrm{disk}$ to drop differs by mega-years, which has an effect on the stellar spin evolution. Furthermore, our model enables the consideration of episodic accretion outbursts \citep[see][]{Steiner21}, based on the layered disk model \citep[][]{gammie96} with $\alpha_\mathrm{MRI} = 0.02$ and $\alpha_\mathrm{DZ} = 0.0004$ \citep[similar to][]{Steiner21}. During such an outburst, the accretion rate can rise for orders of magnitude (yellow line in \fig{fig:alpha}) and the disk's structure and temperature changes significantly. These effects are also calculated in a self-consistent manner. Previous stellar spin evolution studies tend to use a simplified description of the disk's accretion rate. Either an exponential decay \citep[e.g.,][]{Matt10} or a power law description \citep[][]{Gallet19} is used, which require free parameters to define for example the disk lifetime. Additionally, accretion outbursts as well as the back-reactions of such events on the disk cannot be included into these models.

\subsection{Reference model}
\label{sec:refmodel}

Based on the stellar and disk parameter ranges, defined in \sref{sec:parameter_space}, we define a reference model for both a constant $\alpha$-value and a layered disk model.

\subsubsection{Constant $\alpha$-value}

\begin{figure}[ht]
    \centering
         \resizebox{\hsize}{!}{\includegraphics{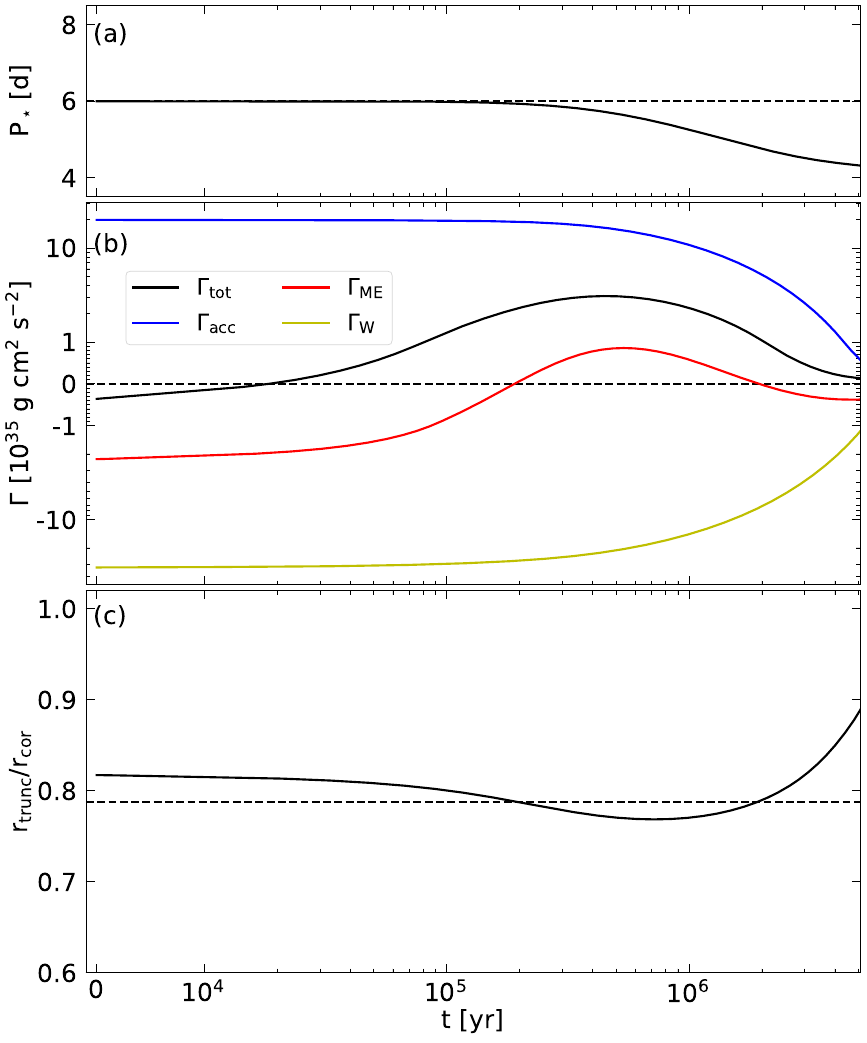}}
    \caption{Stellar spin evolution for \texttt{Mref} with respect to the simulation time $t$ in years starting from $t_\mathrm{0} = 1$~Myr. Panel (a) shows the stellar rotation period $P_\star$ in days. The dashed black line indicates the initial period of 6~days. Panel (b) shows the external torque contributions (see \sref{sec:stellar_spin_model}) as well as the total torque acting on the star (black line). The blue, red and yellow line represent the accretion torque $\Gamma_\mathrm{acc}$, torque due to MEs $\Gamma_\mathrm{ME}$ and the APSW torque $\Gamma_\mathrm{W}$, respectively. The dashed line at $\Gamma = 0.0$ indicate the boundary between a spin up and a spin down contribution. Panel (c) shows the position of the inner disk radius $r_\mathrm{trunc}$ with respect to the position of the co-rotation radius $r_\mathrm{cor}$. The dashed line indicate $K_\mathrm{rot}^{2/3}$.}
    \label{fig:ref}
\end{figure}

The stellar and disk parameters for the reference model using a constant $\alpha$-value (\texttt{Mref}) are summarized in \tab{tab:ref_parameters}. 
The stellar spin evolution for this model is shown in \fig{fig:ref}. During the simulation time of 5~Myr, the stellar rotation period $P_\star$ decreases slightly from the initial value of 6 toward 4~days (see Panel (a)). The individual external contributions to the total torque $\Gamma_\mathrm{tot}$ are shown in Panel (b) (black line). A positive (negative) torque value adds (removes) angular momentum to the star and the star spins up (spins down), respectively. The accretion torque $\Gamma_\mathrm{acc}$ (blue line) spins up the star during the whole simulation time, whereas the APSW ($\Gamma_\mathrm{W}$, yellow line) spins down the star. If MEs ($\Gamma_\mathrm{ME}$, red line) spin the star up or down, depends on the position of the inner disk radius $r_\mathrm{trunc}$ with respect to the co-rotation radius $r_\mathrm{cor}$ (see \equ{eq:gamma_me}), which is shown in Panel (c). For comparison, the value $K_\mathrm{rot}^{2/3}$ is indicated by the dashed black line. 

During the first $2\times10^5$~yr, $r_\mathrm{trunc}$ is close to $r_\mathrm{cor}$ and $r_\mathrm{trunc}/r_\mathrm{cor}>K_\mathrm{rot}^{2/3}$. $\Gamma_\mathrm{ME}$ is negative and the MEs remove angular momentum from the star. The influence of the APSW and the ME, however, are not sufficient to counteract accretion and contraction and the star spins up slightly ($\Gamma_\mathrm{tot}>0$ after $10^4$~years). While the star contracts and the stellar radius as well as the inner disk radius decreases, the co-rotation radius resides at approximately the same positions as the stellar rotational period only changes slightly during the first couple $10^5$~years. Thus, $r_\mathrm{trunc}/r_\mathrm{cor}$ decreases until $r_\mathrm{trunc}/r_\mathrm{cor}<K_\mathrm{rot}^{2/3}$ and the MEs also spin up the star. After $2$~Myr of accretion, the disk is nearly depleted and the disk accretion rates drops (see \fig{fig:alpha}). Consequently, $r_\mathrm{trunc}$ moves outwards and the MEs remove angular momentum from the star again.

\begin{table}[ht]
\centering
\caption{Reference model parameters with a constant $\alpha$-value}              
\begin{tabular}{l r | l r}         
\hline\hline 
$B_\star$ [kG] & 1.00 & $R_\mathrm{\star,~init}$ [$R_\odot$] & 2.096 \\

$P_\mathrm{init}$ [days] & 6.0 & $M_{\star,~\mathrm{init}}$ [$\mathrm{M_\odot}$] & 0.7 \\

$\Dot{M}_\mathrm{init}$ [$\mathrm{M_\odot/yr}$] & $10^{-8}$ & $T_\mathrm{eff,~init}$ [K] & 4078  \\

$W$ [\%] & 2 & $\alpha$ & 0.005 \\

\hline\hline                                            
\end{tabular}
\label{tab:ref_parameters}  
\end{table}

\subsubsection{Layered disk model}

\begin{figure}[ht]
    \centering
         \resizebox{\hsize}{!}{\includegraphics{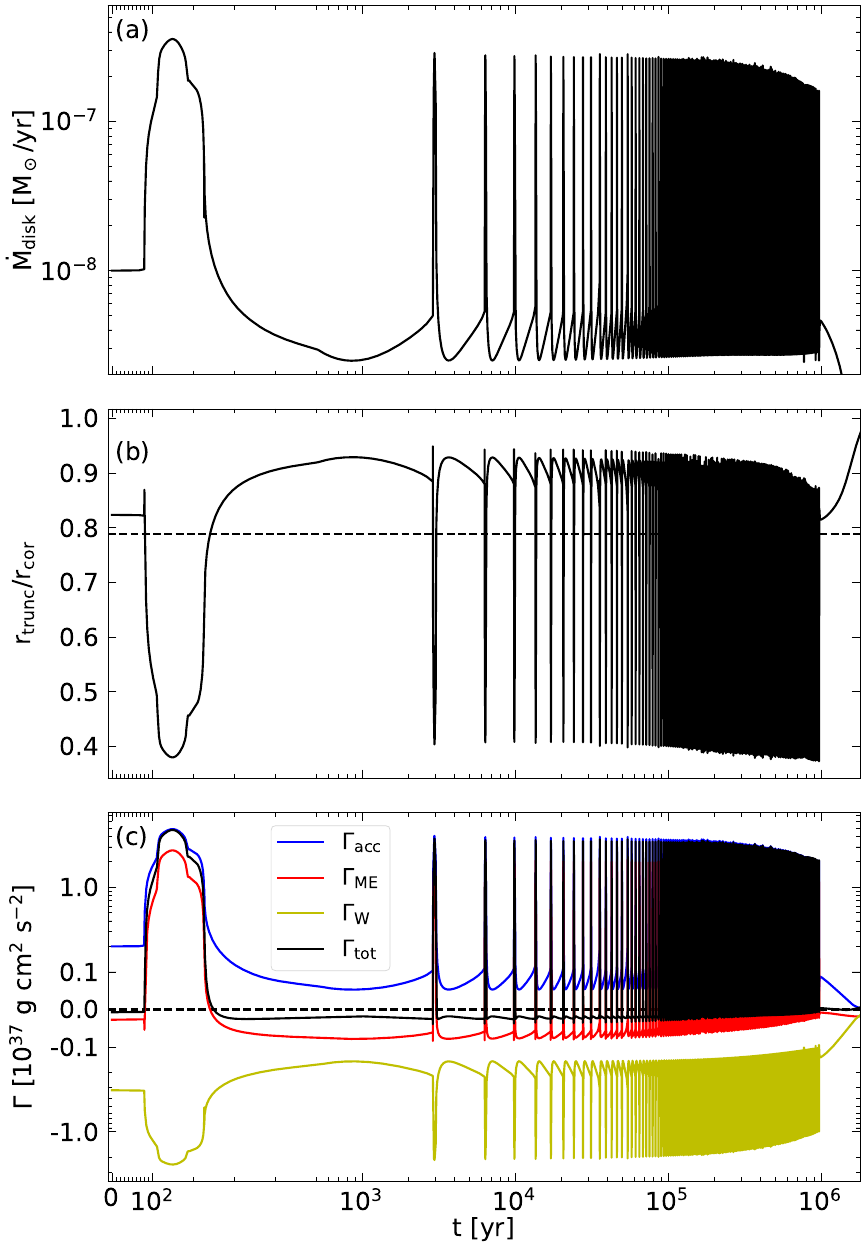}}
    \caption{Stellar spin evolution for \texttt{MrefB} with respect to the simulation time $t$ in years starting from $t_\mathrm{0} = 1$~Myr. Panel (a) shows the disk's accretion rate $\Dot{M}_\mathrm{disk}$. Panel (b) shows the position of the inner disk radius $r_\mathrm{trunc}$ with respect to the positions of the co-rotation radius $r_\mathrm{cor}$. The dashed line indicate $K_\mathrm{rot}^{2/3}$. Panel (c) show the total torque acting on the star $\Gamma_\mathrm{tot}$ as well as the external torque contributions, which are the same as in \fig{fig:ref}. Additionally, the dashed line at $\Gamma = 0.0$ indicate the boundary between a spin up and a spin down contribution.}
    \label{fig:ref_burst}
\end{figure}

The reference parameters (\texttt{MrefB}) used in combination with the layered disk model differ from \texttt{Mref} only by the viscous $\alpha$-value. Instead of a constant $\alpha$-value trough out the disk, we use $\alpha_\mathrm{MRI} = 0.02$ and $\alpha_\mathrm{DZ} = 0.0004$ \citep[similar to][]{Steiner21}. 
The effect of different stellar spin evolutionary tracks based on this reference model are shown in \sref{sec:outbursts}.
In \fig{fig:ref_burst}, a detailed view on accretion bursts and the respective contributions to the total torque acting on the star is shown.
In Panel (a) the disk's accretion rate is shown. Once the disk temperature reaches $T_\mathrm{crit}=1000$~K and the MRI is triggered, the disk's accretion rate $\Dot{M}_\mathrm{disk}$ increases sharply over an order of magnitude until the material piled up in the DZ is depleted and $\Dot{M}_\mathrm{disk}$ decreases again until the next outburst is triggered. 
In Panel (b) the position of the inner radius $r_\mathrm{trunc}$ with respect to the co-rotation radius $r_\mathrm{cor}$ is shown. During the time of increased $\Dot{M}_\mathrm{disk}$, $r_\mathrm{trunc}$ is pushed toward the star (see \equ{eq:truncation_radius}). 
This again effects the total torque acting on the star (black line in Panel (c)). During the first several $10^1$~yr, the total as well as the external torque contributions are similar to the model \texttt{Mref} and the star spins down, which is indicated by $\Gamma_\mathrm{tot}<0$. When the accretion outburst starts, $\Gamma_\mathrm{acc}$ ($\Gamma_\mathrm{W}$) increases (decreases) due to the higher accretion rate, respectively, without changing their sign. 
The contribution of the MEs to $\Gamma_\mathrm{tot}$, however, changes significantly. Before the outburst, $r_\mathrm{trunc}$ is close to $r_\mathrm{cor}$ and $\Gamma_\mathrm{ME}<0$ resulting in a stellar spin down effect. During the burst the distance between $r_\mathrm{trunc}$ and $r_\mathrm{cor}$ increases drastically and $r_\mathrm{trunc}/r_\mathrm{cor}<K_\mathrm{rot}^{2/3}$. 
Now, the MEs spin up the star and the APSW alone cannot counteract the combined spin up torque consisting of $\Gamma_\mathrm{acc}$, $\Gamma_\mathrm{ME}$ and the stellar contraction. As a result, during the accretion outburst, spins up the star. 
While the peak accretion rate of the outburst in \fig{fig:ref_burst} is similar to EXor outbursts, the duration is reminiscent of FUor outbursts \citep[e.g.,][]{Audard14}. 
We note that the peak and duration of these outbursts can be influenced by the choice of $\alpha_\mathrm{MRI}$, $\alpha_\mathrm{DZ}$ and $T_\mathrm{crit}$ \citep[e.g.,][]{Steiner21} and the effects of different outburst types are investigated in appropriate further studies.

\subsection{Fast and slow rotating stars}
\label{sec:fast_slow}

Within the predefined parameter space (\sref{sec:parameter_space}), we define models for a fast and slow rotating star to test the limits of our approach and compare the results to observations of stellar rotation periods of young clusters. Based on our reference model \texttt{Mref}, we change the stellar magnetic field strength $B_\star$, the APSW value $W$ and the initial stellar period $P_\mathrm{init}$.
The fast~(slow) rotation model \texttt{Mfast}~(\texttt{Mslow}) has a stellar magnetic field strength of $B_\star = 500$~(2000)~G, $W$=0~(5)\% and $P_\mathrm{init}=$2~(15)~days, respectively\footnote{
Following previous studies \citep[][]{Matt12, Gallet19}, a weak stellar magnetic field results in a fast rotating star due to the weaker star-disk interaction and a less effective APSW.
}.

\begin{figure}[ht]
    \centering
         \resizebox{\hsize}{!}{\includegraphics{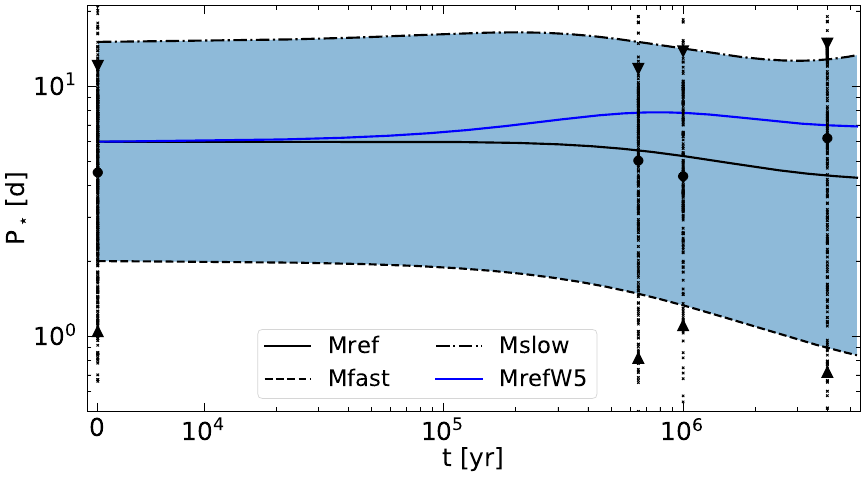}}
    \caption{Evolution of the stellar rotational period $P_\star$ with respect to the simulation time $t$ in years starting from $t_\mathrm{0} = 1$~Myr for the models \texttt{Mref} (solid line), \texttt{Mfast} (dashed line) and \texttt{Mslow} (dash-dotted line).
    Model \texttt{MrefW5} (solid blue line) highlights the dependence on $B_\star$ and the APSW value $W$ (see text).
    The area between \texttt{Mfast} and \texttt{Mslow} is colored for better visualization. The observational data from four young stellar clusters are shown as "crosses" with the 50th percentile (big circle) and the 95th (5th) percentile (downward and upward pointing triangles). From left to right the open clusters are: Orion Nebular Cluster (ONC) \citep[][]{herbst02}, NGC 6530 \citep[][]{Henderson12}, NGC 2264 \citep[][]{Affer13} and NGC 2362 \citep[][]{Irwin08}.}
    \label{fig:spread}
\end{figure}

In \fig{fig:spread}, the rotational periods of \texttt{Mref} (solid line), \texttt{Mfast} (dashed line) and \texttt{Mslow} (dash-dotted line) are shown. 
The stellar rotation periods of four young clusters are also shown to compare the results to the observations. The Orion Nebular Cluster (ONC) with an age of 1~Myr \citep[][]{herbst02} was already used to define the initial parameters used in this study and is located at $t=0$~yr. The following clusters from left to right are NGC 6530 \citep[][]{Henderson12} with an age of 1.65~Myr, NGC 2264 \citep[][]{Affer13} with an age of 2~Myr and NGC 2362 \citep[][]{Irwin08} with an age of 5~Myr. The solid black circles and the downward (upward) pointing triangles represent the 50th and 95th (5th) percentiles, respectively.
The reference model \texttt{Mref} remains close to the 50th percentile value (mean value) of all four clusters. Additionally, the majority of stellar rotation periods in NGC 6530, NGC 2264, as well as NGC 2362 are enclosed between \texttt{Mfast} and \texttt{Mslow}, which is visualized by the blue area between the respective curves.
The spin evolution of \texttt{Mfast} and \texttt{Mslow} depends on the choice of $B_\star$ and $W$\footnote{The influence of the initial stellar period is studied in \sref{sec:back2}.}. 
A weaker (stronger) magnetic field strength result in a shift toward shorter (longer) periods. 
In a similar way, a smaller (larger) APSW value $W$ result in shorter (longer) periods. 
The order of magnitude of this dependence is symbolized by model \texttt{MrefW5} (solid blue line in \fig{fig:spread}). 
Reducing $B_\star$ by a factor of 2 spins up the star by 92\% after 5~Myr (compare models \texttt{Mslow} and \texttt{MrefW5}). 
An increase of $W$ from 2\% to 5\% spins down the star by 60\% after 5~Myr (compare models \texttt{Mref} and \texttt{MrefW5}).

\subsection{Back-reaction on the disk}
\label{sec:backreaction}

In this section, we show the effects stellar rotation has on the protoplanetary accretion disk. First, the effects of different spin evolutionary tracks are analyzed (\sref{sec:back1}), by comparing a star that spins up with a star that spins down. Second, we show the effects that different initial rotation periods have on the accretion disk (\sref{sec:back2}).

\subsubsection{The effect of different stellar spin evolutionary tracks}\label{sec:back1}

To study the effects a star that spins up has on the accretion disk, compared with a star that spins down, we change the APSW efficiency value $W$. While the initial stellar and disk properties are identical the APSW removes more~(less) angular momentum from the star that spins down~(up), respectively. Thus, the effect of different stellar spin evolutionary tracks on the accretion disk, starting from the same initial model, can be shown. 

\begin{figure}[ht]
    \centering
         \resizebox{\hsize}{!}{\includegraphics{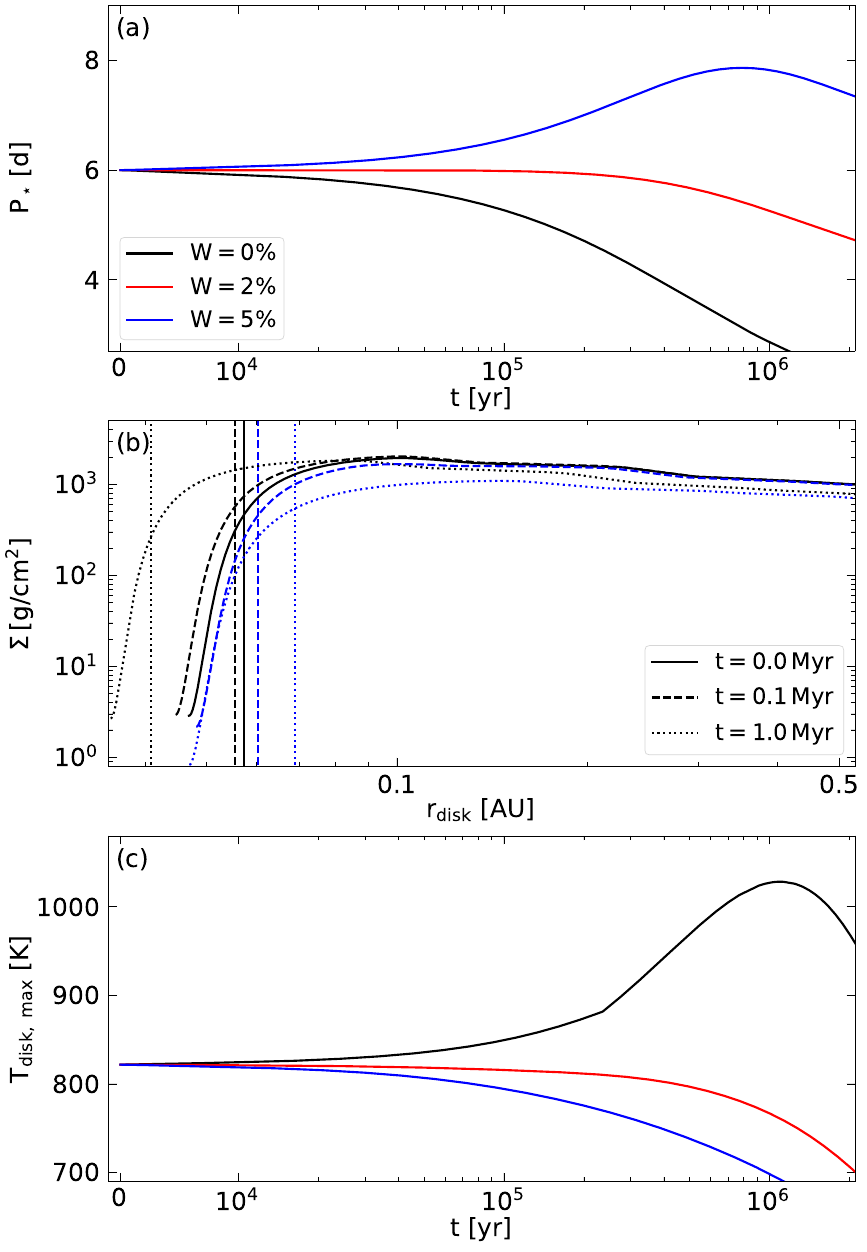}}
    \caption{Panel (a) shows the evolution of the stellar rotational period $P_\star$ with respect to the simulation time $t$ in years starting from $t_\mathrm{0} = 1$~Myr. The reference model \texttt{Mref} with $W=2\%$ , the models with $W=0\%$, and $W=5\%$ are given by the red, black and blue lines, respectively. Panel (b) shows the surface density of the innermost disk part (0.5~AU) for the models with $W=0\%$ (black lines) and $W=5\%$ (blue lines) at three different time snapshots: The initial model at $t=0.0$~Myr (solid line), $t=0.1$~Myr (dashed lines) and $t=1.0$~Myr (dotted lines). The vertical lines represent the position of the co-rotation radius for the given APSW efficiency value and time snapshot. Note that the black and blue solid line are identical as the same initial model ($t=0.0$~Myr) is used for this comparison. Panel (c) shows the evolution of the maximum disk temperature $T_\mathrm{disk,~max}$ at the disk's midplane with respect to the simulation time $t$ in years starting from $t_\mathrm{0} = 1$~Myr. The line colors represent the same models as in Panel (a).}
    \label{fig:accr}
\end{figure}

We choose, based on the reference value of $W=2\%$, $W=0\%$, as well as $W=5\%$ representing a star that spins up and down, respectively. In \fig{fig:accr}, we show the resulting evolution of the stellar rotational period $P_\star$ (Panel (a)). 
The model \texttt{Mref} (red line) evolves as already shown in \fig{fig:ref}. As expected, a low APSW value $W=0\%$ (black line) does not remove enough angular momentum from the star to counteract accretion and contraction and the star spins up during the whole simulation time. 
A large APSW value $W=5\%$ (blue line) removes considerably more stellar angular momentum and the star spins down toward 8~days. 
The resulting effects of these different stellar spin evolutionary tracks on the accretion disk are shown in Panels (b) and (c). 
In Panel (b) the surface density distributions of the innermost disk part ($<0.5$~AU) are shown for different time snapshots: the initial model at $t=0.0$~Myr (solid line), $t=0.1$~Myr (dashed lines) and $t=1.0$~Myr (dotted lines).
Additionally the co-rotation radii are shown for the respective time snapshots (vertical lines). 
Starting from the same initial model (the blue and black solid line are identical), the surface density structure of the inner disk part differs clearly when changing the APSW value from 0 to 5\%. 
Responsible for this difference is the position of the co-rotation radius $r_\mathrm{cor}$. For a star that spins down (blue lines), the co-rotation radius moves outwards away from the star and a larger part of the disk in located inside $r_\mathrm{cor}$. 
Disk material inside $r_\mathrm{cor}$ rotates faster compared to the star and is slowed down due to magnetic torques. 
Thus, the disk's angular velocity ($u_\mathrm{\varphi}$) decreases below the Keplerian velocity and the disk material inside $r_\mathrm{cor}$ accelerates in radial direction toward the star \citep[similar to][]{Steiner21}.
For a star that spins up, $r_\mathrm{cor}$ moves inward toward the star. 
The disk material that is now located outside $r_\mathrm{cor}$ rotates slower compared to the star and is accelerated by magnetic torques to velocities faster than Keplerian. As a result, the radial inward velocity of the disk outside $r_\mathrm{cor}$ is decreased.
Material transported from the outer parts toward the inner disk region, is piled up outside $r_\mathrm{cor}$, pushing the inner disk radius $r_\mathrm{trunc}$ closer to the star.

With the position of $r_\mathrm{cor}$ also the position of the inner disk radius $r_\mathrm{trunc}$ changes. With $r_\mathrm{trunc}$ moving closer to or away from the star, the inner disk is heated differently due to stellar irradiation.
This effect is shown in Panel (c). The maximum disk midplane temperature $T_\mathrm{disk,~max}$ shows a different behaviour depending on the stellar spin evolutionary track. As expected, $T_\mathrm{disk,~max}$ decreases for models that show a stellar spin down (blue and red lines). For the model with $W=0\%$ showing a stellar spin up, $T_\mathrm{disk,~max}$ increases for 1.0~Myr and differs by over 300~K compared to the model with $W=5\%$. Such a difference can have a considerable effect on the disk's ionization fraction that is strongly temperature dependent in these regions of $T_\mathrm{disk,~max}$ \citep[e.g.,][]{Fromang02, Vorobyov20} as well as on the occurrence of episodic accretion outbursts.

\subsubsection{The effect of different initial stellar periods}\label{sec:back2}

\begin{figure}[ht]
    \centering
         \resizebox{\hsize}{!}{\includegraphics{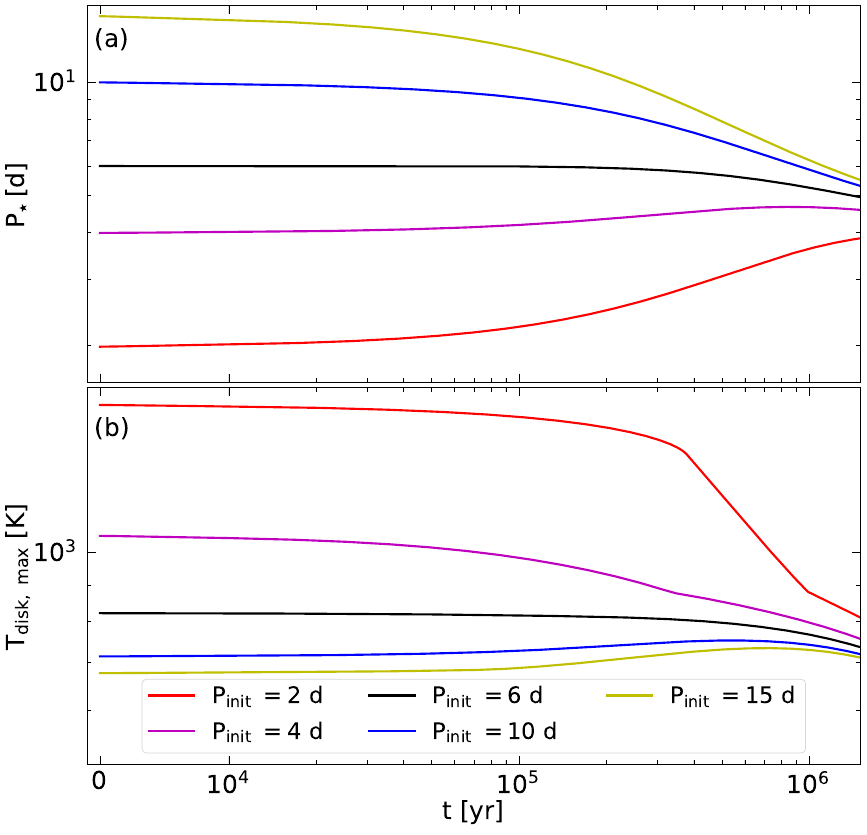}}
    \caption{Evolution of the stellar rotational period $P_\star$ (Panel (a)) and the maximum disk midplane temperature $T_\mathrm{disk,~max}$ (Panel (b)) with respect to the simulation time $t$ in years starting from $t_\mathrm{0} = 1$~Myr. Different colors represent different initial stellar rotational periods: $P_\mathrm{init}=2$~days (red line), $P_\mathrm{init}=4$~days (magenta line), $P_\mathrm{init}=6$~days (black line), $P_\mathrm{init}=10$~days (blue line) and $P_\mathrm{init}=15$~days (yellow line).}
    \label{fig:Pinit}
\end{figure}

When changing the initial stellar rotation period $P_\mathrm{init}$, we expect the period to converge toward a common value in the course of $\sim$~Myr as seen in \cite{Armitage96} or \cite{Matt10}. In \fig{fig:Pinit}, we compare the stellar spin evolution for initial stellar periods ranging from 2 to 15~days. Panel (a) shows the rotation period for $P_\mathrm{init}=2$~days (red line), $P_\mathrm{init}=4$~days (magenta line), $P_\mathrm{init}=6$~days (black line), $P_\mathrm{init}=10$~days (blue line) and $P_\mathrm{init}=15$~days (yellow line). While the stellar periods approach each other, we still see a spread of 2~days after 2~Myr. The reason for this difference is the back-reaction of different stellar spin evolutionary tracks on the accretion disk, which is explained in \sref{sec:back1}. 

The maximum disk midplane temperature $T_\mathrm{disk,~max}$ for each model is shown in Panel (b). Similar to \fig{fig:accr}, a fast rotating star with an inner disk and co-rotation radius close to the star has a higher disk temperature. $T_\mathrm{disk,~max}$ varies by a factor of 2 for the models shown. Again, such a difference strongly affects the disk's ionization as well as influence the occurrence of accretion outbursts.

\subsection{Effect on accretion outbursts}
\label{sec:outbursts}

\begin{figure}[ht]
    \centering
         \resizebox{\hsize}{!}{\includegraphics{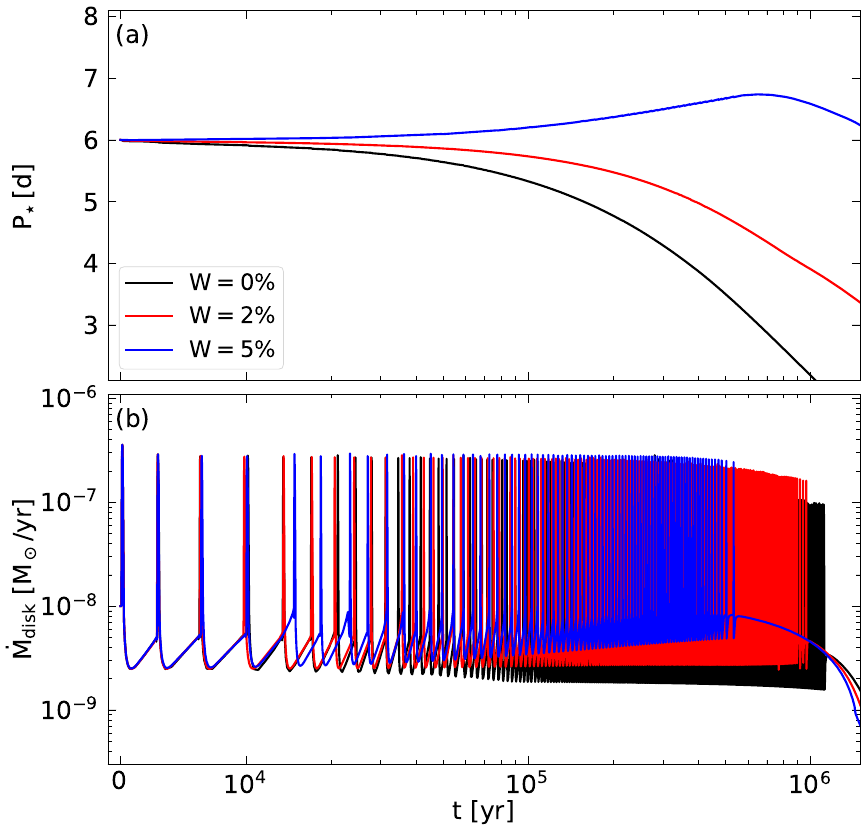}}
    \caption{Stellar spin evolution for different stellar spin evolutionary tracks of the layered disk reference model with respect to the simulation time $t$ in years starting from $t_\mathrm{0} = 1$~Myr. Panel (a) shows the stellar rotational period $P_\star$ for the reference model \texttt{MrefB} with $W=2\%$ (red line) as well as for models with $W=0\%$ (black line) and $W=5\%$ (blue line). Panel (b) shows the disk accretion rate $\Dot{M}_\mathrm{disk}$ for the respective models .}
    \label{fig:burst2}
\end{figure}

Finally, we want to show the effects of different stellar spin evolutionary tracks on the occurrence of accretion outbursts using the layered disk reference model \texttt{MrefB}. In \fig{fig:burst2}, similar to \sref{sec:back1}, different stellar spin evolutionary tracks are shown. Starting from a common initial rotational period of 6~days, the stellar rotational period (Panel (a)) of the reference model \texttt{MrefB} (red line) does not change significantly over a simulation time of 1~Myr, whereas an APSW value of $W=0\%$ results in a stellar spin up (black line) and $W=5\%$ results in a slow rotating star (blue line). In Panel (b), the respective disk accretion rates are shown. Similar to the models in \cite{Steiner21}, outbursts occur over a time span of $\lesssim 1$~Myr. 
The number of bursts as well as the time span on which outbursts are triggered depend on the stellar spin evolution. Fast rotating stars tend to have more mass located closer to the star over longer periods of time (see \fig{fig:accr}). Thus, the disk temperature remains higher within these disks and the conditions that trigger an outburst are also met over longer periods of time. This results in a longer period as well as a larger number of outbursts for stars that spin up and rotate faster. In the model \texttt{MrefB} (red line in \fig{fig:burst2}) 231 outbursts occur over a time span of 0.96~Myr. In the model with $W=5\%$, representing a star that spins down, 85 outbursts occur over a time span of 0.53~Myr, whereas in the model with $W=0\%$, representing a fast rotating star, 835 outbursts occur over a time span of 1.11~Myr. 
Each outburst influences the temperature structure \citep[e.g.,][]{Steiner21} and leads to a change in the chemical composition of the disk \citep[e.g.,][and references therein]{Rab17}. This can change observables such as the snow line \citep[e.g.,][]{Cieza16}, the abundance of $\mathrm{HCO^+}$ \citep[][]{Jorgensen13} or CO observations \citep[][]{Rab17}.

%% \import{sections/}{discussion}

\section{Discussion}
\label{sec:conclusion}

To explain the rotational behaviour of young stars, still surrounded by a protoplanetary disk, the star-disk interaction and its associated torques have to be considered (see \sref{sec:stellar_spin_model}). The disk component in previous studies, however, has been modeled in a simplistic way and important disk features and parameters have not been included in the stellar spin evolution \citep[e.g.,][]{Matt10, Matt12, Gallet19}. To improve the treatment of the disk, we combine hydrodynamic disk and stellar spin evolution to simulate the disk's influence on the stellar rotation and the stellar back-reaction on the disk. The influence of stellar magnetic torques on the disk, the position of the inner disk radius, disk radial density, temperature profiles and the effects on episodic accretion outbursts can be calculated self-consistently during the whole simulation time.

% A combination of the stellar and disk evolution poses some serious problems
A self-consistent treatment of both stellar and disk evolution introduces further complexity. As the stellar spin evolution depends on a precise formulation of the processes of the inner disk, this region has to be well resolved during the whole simulation ranging over several mega-year. Additionally, magnetospheric accretion is an intrinsically multi-dimensional problem \citep[e.g.,][]{Romanova02, bessolaz08, Romanova14}. In addition to timestep limitations imposed by the CFL condition \citep[][]{Courant28}, current explicit, multi-dimensional simulations either exclude the innermost disk regions or treat them in a simplified manner \citep[e.g.,][]{Vorobyov19} or converge toward a steady-state solution within several rotational periods of the inner disk \citep[e.g.,][]{Romanova04, zhu19, Pantolmos20}. These scenarios are not ideal to study the dynamic stellar spin evolution over several mega-year. 
The TAPIR code is able to spatially resolve the inner disk region sufficiently well over the whole disk lifetime. Solving the hydrodynamic equations (cp.~\equs{eq:cont}{eq:ene}), the accretion disk can be modeled in a more self-consistent manner compared to previous stellar spin evolution studies \citep[e.g.,][]{Matt10, Matt12, Gallet19}. We have to note that the one dimensional TAPIR code requires a parameterization of the multi-dimensional processes occurring in the inner disk. Consequently, the results presented above are based on simplifications. Nevertheless, the general trends and dependencies shown in this study should be valid.

Young stars are assumed to have strong magnetic fields \citep[e.g.,][]{Krull09, Gregory12, Johnstone14}. The complex field topology and temporal variability of the field, however, is difficult to observe or to predict and equally challenging to develop in a numerical model, which describes the field dynamics correctly and simultaneously maintains the ability to carry out long-term simulations.
Based on the measurement technique, different values for the stellar magnetic field are observed for the same star. This issue can be explained on the basis of the young M dwarf AU~Mic \citep[][]{Kochukhov20}. A mean magnetic field of $B_\mathrm{mean} = 2.3$~kG results from Stokes
I profile measurements. With a more detailed Zeeman Doppler imaging technique based on the Stokes V parameter, a global field strength of $B_\mathrm{global} = 0.09$~kG is observed. Finally, based on the Stokes Q and U parameters a dipole field strength of $B_\mathrm{dipole}=2.0$~kG is obtained. Thus, the observed magnetic field can change based on the used measurement technique by a factor of 20.
\cite{Finnley18}, however, have shown in stellar wind calculations that usually the lowest order field component dominates $\Gamma_\mathrm{w}$.
Additionally, stellar magnetic fields can vary on short timescales. \cite{Johnstone14} have shown that the dipole magnetic field strength of V2129 Oph varies by a factor of almost 4 over a four year period.

Stellar winds, in this study represented by APSW, strongly affect the long-term stellar spin evolution. While an APSW influences not only the stellar spin evolution providing a possible solution to the occurrence of slow spinning stars, it also affects the disk structure as well as its long term evolution (see \fig{fig:accr} and \fig{fig:burst2}). The exact nature and strength of these winds, however, are still debated \citep[see the discussion in][]{Gallet19}. 
Previous authors need large APSW values of $W \gtrsim 10\%$ to explain slowly rotating stars \citep[e.g.,][]{Matt12,Gallet19, Ireland21}, which contradicts the results of \cite{Crammer08} and \cite{Pantolmos20}, limiting $W\sim 1\%$.
We can reduce this value to $W\lesssim 5\%$ to reproduce the observed rotational period of most stars in young stellar clusters (see \fig{fig:spread}).
Apart from APSW, other wind mechanisms can be important for the stellar and disk evolution. Especially magnetic disk winds affect the disk evolution noticeably \citep[e.g.,][]{Konigl11}. By removing matter and, in the presence of a large-scale magnetic field threading the disk, angular momentum from the disk before it can be accreted onto the star, the stellar spin evolution is affected. 
Similar, photo-evaporation (internal or external) can influence the disks lifetime and thus, the stellar spin evolution \citep[e.g.,][]{Roquette21}.
These effects are added to our model in further studies.

The stellar spin evolution is strongly influenced by the initial stellar parameters. Besides the magnetic field strength, the initial radius as well as the temperature play an important role. The change of the stellar moment of inertia ($\Dot{I}_\star$) can vary significantly based on the choice of initial parameters (\equo{eq:stellar_radius_evolution}).

We take initial stellar values from \cite{Baraffe15} and evolve the star using a rather simplified model to describe $\Dot{R}_\star$. As described in \cite{Matt10} such assumptions reproduce more sophisticated stellar evolution models to a reasonable degree of accuracy. \cite{Siess00}, however, concludes that the variation of, for example, stellar metallicity can vary $T_\star$ by $\gtrsim 200$~K.
Additionally, a spin up or spin down torque can alter the rotational period and thus stellar evolution \citep[][]{Zeipel24, Endal76, Maeder00, Tassou00, Praxton19, Amard19}. Moreover, the amount of accreting energy that is radiated away from the star influences the stellar evolution. The deviation in stellar radius and luminosity between models, in which most of the accreted energy is radiated away (cold accretion) and a certain amount of accreted energy is added to the star (hybrid or hot accretion), can reach up to 50\% \citep[e.g.,][]{Baraffe09, Vorobyov17c}. 
Especially during an accretion outburst, the stellar radius and effective temperature is influenced \citep[e.g.,][]{Vorobyov17c}, which is not yet included into our model.
Thus, the combination of our disk hydrodynamics model with a more elaborate stellar evolution code \citep[e.g., the MESA code;][]{Praxton19} could include these effects in a self-consistent way \citep[similar to e.g.,][]{Vorobyov17d} and would be a further improvement of the stellar spin evolution model.

\section{Summary and conclusion}
\label{sec:conclusion2}

In this study, we have enhanced the hydrodynamic disk evolution TAPIR code \citep[][]{ragossnig20, Steiner21} with a stellar spin evolution model (see \sref{sec:stellar_spin_model}). Several properties of the disk (e.g., the disk's accretion rate or the disk lifetime) that have been approximated in previous studies \citep[e.g.,][]{Matt10, Gallet19} are now calculated in a self-consistent way. 
Additionally, the back reaction of stellar rotation on the disk can be included. Different stellar spin evolutionary tracks and the associated movement of the co-rotation radius influences the inner disk structure. The inner disk radius of a star that spins up is located closer to the star (see \fig{fig:accr}). More disk material is heated up due to the proximity to the star and the disk temperature increases. Similarly, the disk around a star with large initial rotational period remains relatively cold, compared to a initially fast rotating star (see \fig{fig:Pinit}). 
From our range of parameters presented in \sref{sec:parameter_space}, we can produce fast and slow rotating models (see \fig{fig:spread}), which are able to enclose most of the stellar rotational periods observed in young stellar clusters \citep[e.g.,][]{Henderson12, Gallet13} with smaller APSW values compared to previous studies \citep[e.g.,][]{Matt12, Gallet19, Ireland21}. 
Furthermore, for the first time, the effects of episodic outburst events on the stellar rotation and vice versa can be studied over several mega-year. Disks around stars that spin up show a larger number of outbursts as well as a longer outbursting period, compared to slowly rotating stars.
While we have shown the effects of different stellar spin evolutionary tracks on the occurrence of outbursts, further studies have to be conducted to analyze a wider range of parameters such as, for example, different stellar and disk masses, outburst strengths as well as including the early evolution of the star-disk system (setting $t_\mathrm{0}$ to $\lesssim 10^4$~years). 
Our model that combines the stellar spin and hydrodynamic disk evolution enables us to include these effects in further studies and work toward a better understanding of the stellar spin evolution.

\begin{acknowledgements}
The authors are thankful to the anonymous referee for useful suggestions that helped to improve the manuscript. 
E. I. V. acknowledges support of Ministry of Science and Higher Education of the Russian Federation under the grant 075-15-2020-780 (N13.1902.21.0039; Sect. 3 and 4).
\end{acknowledgements}

% ------------------------------------------------------------------
% FOOTER
\bibliographystyle{resources/bibtex/aa}
\bibliography{literature/tapir}

\end{document}